\documentclass[useAMS,usenatbib]{mnras}
 \usepackage{setspace}
\usepackage{graphicx}
\usepackage{natbib}
\usepackage{color,ulem}
\title[Star Formation \& Metals in an LSB Galaxy]{The Distribution of Star Formation and Metals in the Low Surface Brightness Galaxy UGC 628}

\author[J. E. Young, R. Kuzio de Naray, and Sharon X. Wang]
  {J.~E.~Young,$^1$
   Rachel Kuzio de Naray,$^2$
   Sharon X. Wang$^3$\\
  $^1$Department of Science, Mount St. Mary's University, 16300 Old Emmitsburg Road, Emmitsburg MD, 21727 jyoung@msmary.edu\\
  $^2$Department of Physics \& Astronomy, Georgia State University, P.O.\ Box 5060, Atlanta, GA 30302-5060 kuzio@astro.gsu.edu\\
  $^3$Department of Astronomy \& Astrophysics, The Pennsylvania State University, 525 Davey Lab, University Park, PA 16802 xxw131@psu.edu\\\\\large \textnormal{Accepted for publication in MNRAS}}

\date{June 2015}

\pagerange{\pageref{firstpage}--\pageref{lastpage}} \pubyear{2015}

\def\LaTeX{L\kern-.36em\raise.3ex\hbox{a}\kern-.15em
    T\kern-.1667em\lower.7ex\hbox{E}\kern-.125emX}

\newcommand{\changed}{\textcolor{red}}
\renewcommand{\changed}{}

\begin{document}

\label{firstpage}

\maketitle

\begin{abstract}
  We introduce the MUSCEL Program (MUltiwavelength observations of the Structure, Chemistry and Evolution of LSB galaxies), a project aimed at determining the star-formation histories of low surface brightness galaxies. MUSCEL utilizes ground-based optical spectra and space-based UV and IR photometry to fully constrain the star-formation histories of our targets with the aim of shedding light on the processes that led low surface brightness galaxies down a different evolutionary path from that followed by high surface brightness galaxies, such as our Milky Way. Here we present the spatially-resolved optical spectra of UGC~628, observed with the VIRUS-P IFU at the 2.7-m Harlen J.\ Smith Telescope at the McDonald Observatory, and utilize emission-line diagnostics to determine the rate and distribution of star formation as well as the gas-phase metallicity and metallicity gradient. We find highly clustered star formation throughout UGC~628, excluding the core regions, and a log(O/H) metallicity around $-4.2$, with more metal rich regions near the edges of the galactic disk. Based on the emission-line diagnostics alone, the current mode of star formation, slow and concentrated in the outer disk, appears to have dominated for quite some time, although there are clear signs of a much older stellar population formed in a more standard inside-out fashion.

\end{abstract}


\def \arcsecond#1#2{$#1''\!\!.#2$}
\def \arcminute#1#2{$#1'\!\!.#2$}
\def \micron{$\mu$m }
\def \angstrom{$\rm\AA\;$}
\def \magsq{$\rm mag''$}
\def \deg{$^\circ$}

\makeatletter
\newcommand{\rom}[1]{\romannumeral #1}
\newcommand{\Rom}[1]{\expandafter\@slowromancap\romannumeral #1@}
\makeatother

\newcommand{\ionl}[3]{[#1$\,${\small \Rom{#2}}]\relax$\lambda$#3}

\def\Pa{{\rm Pa}$\alpha$}
\def\HA{{\rm H}$\alpha$}
\def\HB{{\rm H}$\beta$}
\def\HG{{\rm H}$\gamma$}
\def\HD{{\rm H}$\delta$}
\def\HE{{\rm H}$\varepsilon$}
\newcommand{\HI}{H{$\,$\footnotesize\Rom{1}}}
\newcommand{\HII}{H$\,${\small II}}
\newcommand{\AV}{A$_{\rm V}$}

\newcommand\nobrkhyph{\mbox{-}}
\newcommand\vs{versus}
\newcommand\rfrac[2]{${}^{#1}\!/_{#2}$}

\def\apj{ApJ}
\def\aap{A\&A}
\def\apjl{ApJ}
\def\aj{AJ}
\def\araa{ARA\&A}
\def\mnras{MNRAS}
\def\nat{Nature}
\def\apjs{ApJS}
\def\jrasc{JRASC}

\section{Introduction}
\label{sec:intro}

Low surface brightness (LSB) spiral galaxies represent a significant gap in our understanding of star formation: They are gas-rich disks which span the same range of baryonic masses as their high surface brightness (HSB) counterparts, yet, for reasons not fully understood, they have largely retained their atomic gas instead of producing stars. The difficulty in detecting LSB galaxies makes their cosmological significance challenging to estimate; studies based on optical surveys \citep{McGaugh1995,ONeil2000,Trachternach2006} indicate that LSB galaxies account for up to 50\% of the galaxy-bound baryons in the universe, while \cite{Hayward2005} use supernovae to estimate that fraction closer to 10\%.

Attempts to place LSB galaxies within the context of hierarchical formation using broad-band colors have been hampered by their paradoxical properties, such as blue colors but faint stellar populations and low star-formation rates despite gas rich disks. The only universal conclusion that has been reached is that the star-formation histories of LSB galaxies differ from their HSB counterparts.

Among the many descriptions of LSBs, a favored phenomenological explanation for the blue colors and low star-formation rates is that LSB galaxies follow a burst-and-quench cycle, where small regions sporadically erupt with short bursts of star formation just frequently enough to keep the galactic disk blue without building up a large stellar population \citep{Zackrisson2005,Vorobyov}. Observations of patchiness in \HA{} images of LSB galaxies \cite[e.g.,][]{Auld2006,Kim2007}, consistent with the brief but concentrated star formation proposed by \cite{Vorobyov}, support this hypothesis.

The burst-and-quench scenario is in line with a significant clue from the ubiquitously low gas surface density in LSB galaxies (despite a globally high $\rm M_{gas}/M_\star$ ratio), with $\rho_{\rm gas}$ rarely climbing above the Kennicutt-Schmidt threshold \citep{KennicuttSchmidt1998}. Several authors \citep[e.g.,][]{deBlok1996,Mihos1997,Schombert2013} argue that the gas density is too low to allow for self-gravitating global patterns, leaving star formation (such as it is) to be governed by local fluctuations. For example, while 50\% of HSB galaxies host nuclear star formation, \cite{Schombert2013} found that nuclear star formation is extremely rare in their sample of 54 LSB galaxies. Likewise, \cite{Mihos1997} point out that barred LSB spirals appear to make up 3-4\% of several samples of LSB galaxies \citep{McGaugh1994,Bothun1995,Impey1996}, while 30\% of isolated HSB spirals have bars \citep{Elmegreen1990}.

The role of global patterns, however, is far from clear, as \cite{Kim2007} found that in a sample of 13 spiral and/or barred LSB galaxies seven show centralized star formation and six show correlation between star formation and the arms and/or bars, suggesting that global patterns may be significant. Global patterns may also be more of a side effect than the ultimate cause; several authors \citep[e.g.,][]{Bothun1993,Mo1994} have hypothesized that LSB galaxies have low star-formation rates because they are more isolated, and interactions are a major source of galaxy-wide patterns. Supporting this idea, \cite{Rosenbaum2004} found that LSB galaxies in SDSS data are much more likely to be found on the edges of large scale structure, likely having formed in the voids.

Another significant clue to the star-formation histories of LSB galaxies may be found in their dust and metal content. With modest stellar populations and high gas fractions LSB galaxies are generally assumed to be largely unevolved and relatively free of dust and metals. This assumption is circumstantially supported by weak or absent thermal dust emission in the mid- and far-IR \cite{Hinz2007}, however, since thermal dust is more steeply correlated with UV heating than it is with actual dust content, the absence of thermal dust emission is likely linked to the low star-formation rate. Indeed, \cite{Hinz2007} note that in their sample of five LSB galaxies thermal dust emission is found only in those with \HII{} regions.

In a comprehensive study \cite{McGaugh1994abundance} spectroscopically surveyed \HII{} regions in 22 LSB galaxies, and found that the reddening and log(O/H) metallicities span a wide range, albeit with averages lower than HSB galaxies. \cite{Vorobyov} point out that the burst-and-quench scenario leaves the signature of metallicity variations across the galactic disk {\it if} LSB galaxies are truly unevolved; since no such fluctuations are seen in observations, LSB galaxies appear to have been forming stars for quite some time, albeit very slowly.

To thoroughly address the star-formation histories of LSB galaxies, we have undertaken a multiwavelength study of a sample of gas-rich Milky Way mass LSB spirals centered around optical IFU observations; this is the first in a series of papers which will closely examine current and past star formation in LSB galaxies with the aim of placing this difficult to observe galaxy class in the context of galaxy formation and shedding light on the physics which regulate star formation.

In this paper we examine the optical emission-line strengths and ratios of \HII{} regions in our first target, UGC~628. The information available via the emission lines allows us to place this LSB galaxy in the context of the various models and scenarios proposed, and will supplement the results in an upcoming series of papers. Specifically, we examine the optical emission-line properties of UGC~628, focusing on the \ionl{O}{2}{3727}, \HB{}, \ionl{O}{3}{4959}, and \ionl{O}{3}{5007} lines. These lines, particularly the \HB{} line, allow us to address the rate and distribution of current star formation in UGC~628, while the line ratios act as windows into the star-formation history in UGC~628.

In Section \ref{sec:observations} we describe our target selection, observations, and basic data reduction, in Section \ref{sec:analysis} we describe the methodology used to process the reduced data into narrow-band images, in Section \ref{sec:results} we compare our findings to those in other studies of LSB and HSB galaxies, and in Section \ref{sec:discussion} we summarize our findings and place UGC~628 in the context of galaxy evolution.

\section{Observations}
\label{sec:observations}

Our program, MUSCEL (MUltiwavelength observations of the Structure, Chemistry and Evolution of LSB galaxies), is designed to unambiguously determine the star-formation histories of LSB galaxies, a goal which faces the significant challenge that the parameters of model galaxies are often degenerate over observations in any one wavelength regime. To combat this issue, we have designed our program in three segments: space-based UV observations (mostly via the Swift UVOT instrument), which fully constrain the current star-formation rate; archival Spitzer IRAC observations, which fully constrain the integrated star-formation history; and ground-based optical spectra, which are sensitive to intermediate age indicators, such as the shape of the stellar continuum, Balmer absorption lines, and the 4000\angstrom break.

In order to match the $\sim$2\arcsec{} angular resolution of our broad band UV and IR images, we chose to make our spectral observations with the VIRUS-P integral field unit (IFU), a multi-fiber spectrograph on the 2.7-m Harlen J.\ Smith Telescope at Mt.\ Locke, which has the capability to provide a spectrum at every location in the field of view, essentially giving us an image of our target at every wavelength within the range of our instrument. The VIRUS-P wavelength range encompasses several bright optical emission lines which, by themselves, provide significant clues to the star-formation history of our first target, UGC~628. An upcoming paper will examine the star-formation history fitted against the full UV/optical/IR SED, but here we present an analysis of the emission lines alone.

\subsection{Target Selection and Background}
We chose targets drawn from the LSB galaxies cataloged in \cite{Kim2007} and \cite{McGaugh1994} that fit comfortably within the \arcminute{1}{7}$\times$\arcminute{1}{7} VIRUS-P field-of-view while still spanning several 2\arcsec{} seeing-limited resolution elements (typical of Mt. Locke). To maximize galaxy coverage, minimize dust obscuration, and minimize the number of distinct stellar populations along each line-of-sight, we avoided galaxies that are edge-on ($i>85^\circ$). Additionally, since the primary objective of our program is an evaluation of candidate star-formation histories, we restricted our targets to LSB galaxies with archival Spitzer IRAC observations to constrain the mature stellar population. The first target in our program with completed ground-based spectra is UGC~628, which is the focus of this paper.

With a central B-band surface brightness of $\mu_0=23.1\,\rm mag\, arcsec^{-2}$ \citep{Kim2007} UGC~628 falls well within the LSB category, defined as $\mu_0 > 22$ by \cite{McGaugh1995}, though it is by no means an extreme member of the population $(\mu_0 > 24)$. Using broad-band photometry \cite{Kim2007} estimate $\rm log(M_\star/M_\odot) =$ 10.65 or 10.80, depending on the IMF assumed, making it a near match to the $\rm log(M_\star/M_\odot)=10.7$ for the Milky Way \citep{Flynn2006,McMillan2011}.

\begin{figure}
\begin{center}
  \includegraphics[width=\linewidth]{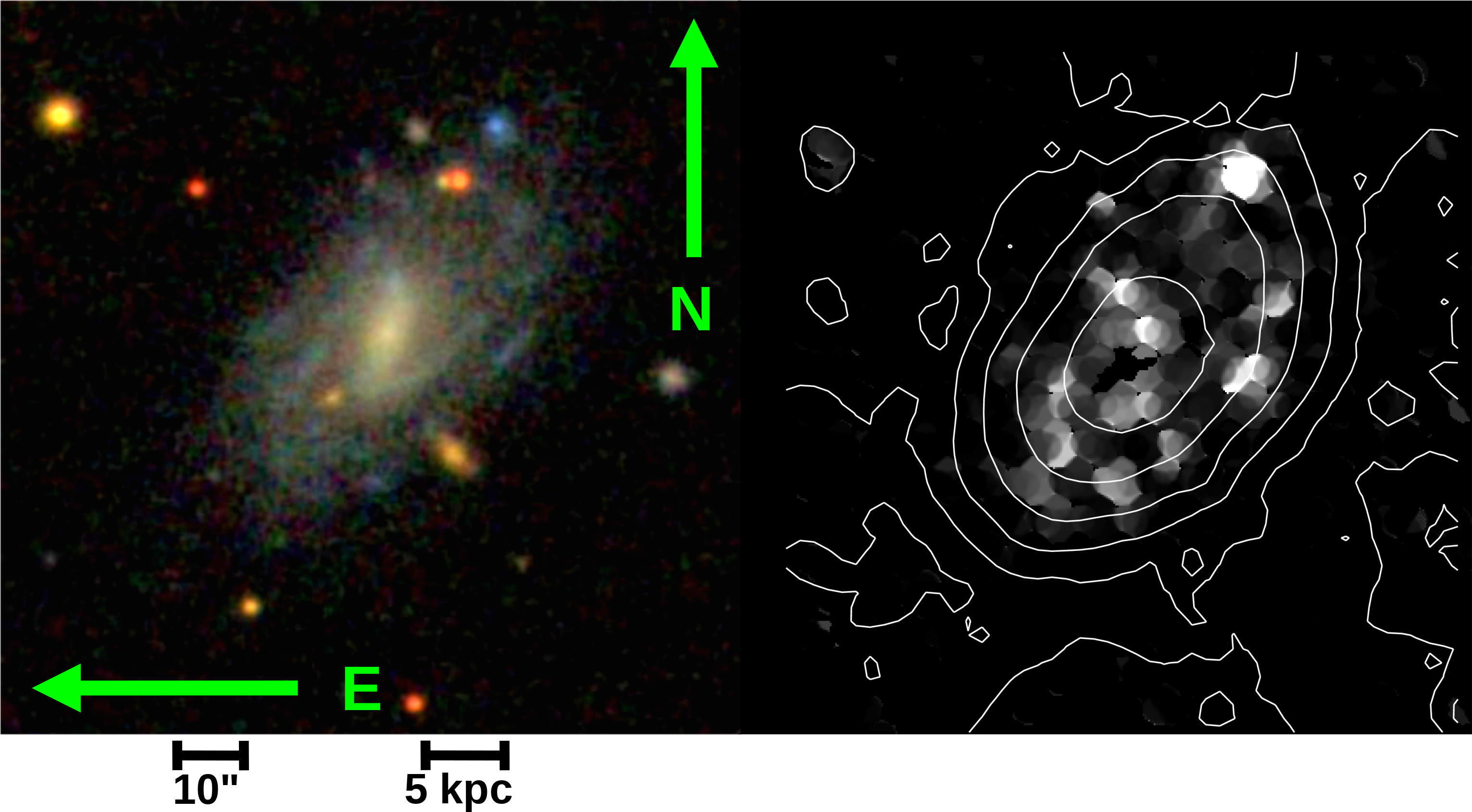}
  \caption{
    {\bf(a, left)} SDSS image of UGC~628. 
    {\bf(b, right)} An \HB{} image of UGC~628 overlayed with stellar continuum isophotes.
  }
\label{fig:spitzerRGB}
\end{center}
\end{figure}

UGC~628 is an extreme late-type spiral (see Figure~\ref{fig:spitzerRGB}(a)), classified as a Magellanic Spiral (Sm) by \cite{RC3}. With a redshift of $z\!=\!0.018166$ \citep{Schneider1990}, the proper distance to UGC~628 is approximately 78Mpc. Like many LSB galaxies, UGC~628 is dark matter dominated, with nearly solid-body rotation out to $\sim$3 kpc \citep{deBlokBosma2002}, but, unlike most LSB disks, UGC~628 is a barred spiral. \cite{Mihos1997} point out that only a few percent of LSB disks show bars, compared to the 30\% of HSB spirals reported in the RC2 catalog \citep{Elmegreen1990}. 

The bar in UGC~628 is itself also extremely unusual in that it is a slow bar, meaning that bar pattern speed is slow enough that the corotation radius is more than 1.4 times the semi-bar length \citep{Chemin2009slowbar}. This is one of only three slow bars known, the others being a stellar bar in the dwarf irregular galaxy NGC~3741 \citep{Banerjee2013} and a purely gaseous slow bar in the dwarf irregular galaxy NGC~2915 \citep{Bureau1999}.

\subsection{Data Acquisition}
\label{sec:acquisition}
We collected our data during a four night observing run in November 2013. The weather was mostly clear for one and one half of these nights, and during this time we completed observations of UGC~628. 

The VIRUS-P IFU consists of 246 fibers permanently packed into a hexagonal bundle and placed at the focal plane with the spectrally dispersed images of the fibers refocused onto a 2048$\times$2048 CCD. Since the aim of our project is to determine the shape of the stellar continuum in our target galaxies, we used the VP1 grating with a 5.1\angstrom resolution to maximize wavelength coverage, spanning roughly 3525\angstrom to 5775\angstrom.

We planned our observations in a six-point dither pattern. The purpose of the six-point dither pattern is two-fold. First, because the VIRUS-P IFU has a \rfrac{1}{3} fill factor, a minimum of three pointings is needed to completely fill the field-of-view. Second, since \arcsecond{4}{16} VIRUS-P fibers undersample the 2$''$ seeing (typical for Mt. Locke), over dithering by a factor of two is necessary to approach the seeing-imposed limit (see Section \ref{sec:weave}).

In addition to the VIRUS-P IFU frames, we took contemporaneous frames with the fixed-offset guide camera. The guide camera data were used to calibrate the relative astrometry of the IFU data, which enabled reconstruction of the field-of-view (see Section \ref{sec:weave}).

\subsection{Basic Reduction}
\label{sec:basic}
In addition to the science frames we also collected bias, flatfield, and wavelength calibration data at the beginning and end of each night. We median-combined all of the bias frames taken on each of the four nights to create a nightly master bias frame, with which we removed the bias from each of the non-bias frames on each respective night. The remaining flat-fields, target frames, sky frames, and wavelength calibration frames were then cleaned to eliminate cosmic rays. 

The cleaned wavelength comparison frames were median-combined to produce high S/N cosmic-ray-free master wavelength comparison frames for each evening and morning, which we then compared to a Hg-Cd line list to create an evening and morning wavelength solution for each night.

Using this wavelength solution the cleaned flat, target, and sky frames were then wavelength-rectified and collapsed --- that is, within each $x$ column all of the pixels associated with each fiber were summed to a single value, and those values were binned with regular steps in wavelength to align the fibers. To account for variations in the telescope and spectrograph over the course of the night, we used the wavelength solution (evening vs. morning) taken closest in time to each respective frame.

Similarly, to create a master flat-field for the evening and morning of each night, we normalized and median combined the collapsed twilight flat-fields. All 246 spectra in each sky flat were divided by the underlying sky spectrum, which we determined by taking the median value of the 246 spectra. Each of the target and sky frames were flattened and collapsed using the flat field, evening or morning, that was closest in time.

\subsection{Sky Subtraction}
\label{sec:skysub}
The surface brightnesses of our program's targets are all fainter than the background of even the darkest skies, making precise sky subtraction paramount. Fortunately, our targets are all smaller than the VIRUS-P field-of-view, leaving the fibers at the edge of the field-of-view as contemporaneous sky measurements. In our final data products, we adopt the median of the faintest forty fibers in each target exposure as the sky background spectrum, and subtract this background from the remaining fibers. Our algorithm did not make any a~priori assumptions about the location of the forty sky fibers, but we confirmed after the fact that the selected fibers were indeed at the edges of the IFU, thus avoiding any galaxy light which could have otherwise affected our sky calculation.

Because the surface brightness in some regions of our target is more than 10$\times$ fainter than the sky background, uncertainties in sky subtraction dominate the uncertainties in our flux measurements. Therefore, we adopt the standard deviation of the fluxes of the faintest forty fibers as the uncertainty in the sky background and propagate it forward in parallel with the flux data.

\begin{table}
\begin{center}
\caption{Comparison to Broad-Band Photometry}
\begin{tabular}{lccc}
\hline
Source
&U
&B
&V
\\
\hline

This Work       & 15.4$\pm$0.2 & 15.66$\pm$0.07 & 15.02$\pm$0.05 \\
\cite{Kim2007}  & 15.5$\pm$0.1 & 15.6$\pm$0.1   & 15.1$\pm$0.1 \\
\hline

\label{tbl:compare}
\end{tabular}
\end{center}
\end{table}

\subsection{Field-of-View Reconstruction}
\label{sec:weave}
To reconstruct the field-of-view at each wavelength we must take into account the relative position of each fiber with respect to the other fibers (provided by the McDonald Observatory) as well as the sky position of the IFU in each exposure (provided by the guide camera astrometry). For the purposes of our project, we mapped the spectral information from every fiber in every dither into an $x$-$y$-$\lambda$ spectral cube with a \arcsecond{0}{3}$\times$\arcsecond{0}{3} plate scale and a 3\angstrom wavelength scale. This plate scale oversamples the 2$''$ resolution imposed by typical seeing at Mt. Locke. It is, nonetheless, quite appropriate since a plate scale closer to 2$''$ would badly undersample some of the resolution elements because the fibers are circular (as opposed to square) and arranged in a hexagonal configuration (as opposed to a square grid). Likewise, we have chosen a 3\angstrom wavelength scale, oversampling the 5.1\angstrom resolution by almost a factor of two. A choice closer to 5.1\angstrom would effectively blur out by almost a factor of two narrow spectral features straddle pixel boundaries.

We used a weight function for each fiber that represents the fraction of the exposure time during which that fiber illuminated a given location within the image plane. For each exposure we determined the IFU position and drift radius from the guide camera images taken during that exposure. Our guide camera data as well as the VIRUS-P handbook indicate that the instrument drifts by roughly \arcsecond{0}{5} during exposures. Each pixel in the reconstructed $x$-$y$-$\lambda$ spectral cube is a thus weighted average of all calibrated IFU spectra from all of the exposures of that target. The calibrated $x$-$y$-$\lambda$ spectral cubes are our primary data products, encoding the flux at every location within our targets at every wavelength covered by VIRUS-P. As described in Section \ref{sec:skysub}, the uncertainties in the $x$-$y$-$\lambda$ spectral cube are dominated by uncertainties in the sky subtraction, typically around a~few~$\times10^{-19} \rm erg\;s^{-1} cm^{-2} \AA^{-1}arcsec^{-2}$.

\begin{figure}
\begin{center}
  \includegraphics[width=\linewidth]{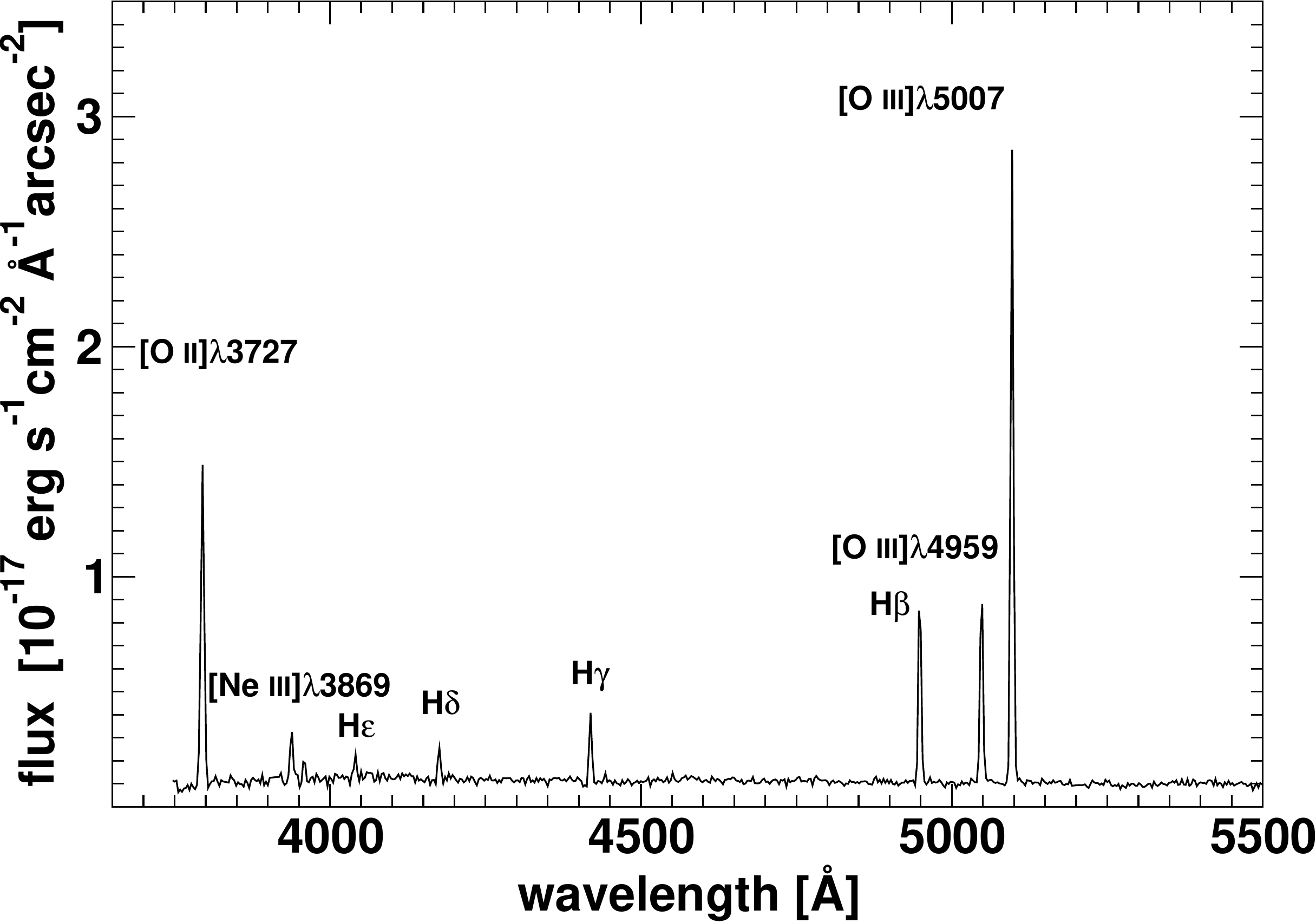}
  \caption
    {Spectrum of region $A^\prime$ (see Section \ref{sec:regions}) with bright emission lines marked; the other \HII{} regions in UGC~628 show similar emission line characteristics, albeit at lower S/N.}
\label{fig:spectrum}
\end{center}
\end{figure}

\subsection{Spectrophotometric Calibration}
\label{sec:calib}

To correctly calibrate our target data we observed several standard stars at different airmasses each night. In order to ensure that the standard star flux is not truncated by the  \arcsecond{4}{16} fibers or the \rfrac{1}{3} fill factor of the VIRUS-P IFU, we compared the reference spectra to aperture photometry from a reconstructed $x$-$y$-$\lambda$ spectral map of the field-of-view. In cases where the standard star's field-of-view was not completely covered, we reconstructed it using the azimuthal symmetry of the PSF.

As a final step in our calibration, we corrected the fluxes in each wavelength bin in our $x$-$y$-$\lambda$ spectral maps for foreground galactic extinction using the Cardelli extinction law \citep{Cardelli} and E(B-V)=0.0376, derived from the dust maps in \cite{Schlegel1998}.

As a check on our calibration, we passed the VIRUS-P data for UGC~628 through UBV filter transmission curves to recreate a UBV image. We compare the whole-galaxy UBV photometry to established values \citep{Kim2007} in Table~\ref{tbl:compare}; the photometry in all three bands matches to within 0.1 magnitudes.

\section{Analysis}
\label{sec:analysis}

\subsection{Emission Line Images}
\label{sec:lineimages}
In Figure~\ref{fig:spectrum} we present the spectrum from a region of UGC~628 particularly bright in line-emission. In order to identify regions with bright line emission and take advantage of the information available from line ratio diagnostics, we created emission-line-only slices from the $x$-$y$-$\lambda$ spectral cube centered on the \ionl{O}{2}{3727}, \HG{}, \HB{}, \ionl{O}{3}{4959}, and \ionl{O}{3}{5007} lines. Each slice is the sum of all the flux within an 11\angstrom filter centered on each emission line minus half the average of the sums of two 22\angstrom$\!\!$-wide continuum slices, each located 6.5\angstrom away from the emission line. The 11\angstrom$\!\!$-wide filter generously includes all of the 5.1\angstrom PSF, and the flanking continuum filters effectively take into account any local slope in the stellar continuum. Additionally, we account for the stellar absorption underneath the Balmer emission lines by adding 2\angstrom of continuum flux back into the emission lines.

This method of accounting for the stellar absorption underneath the emission lines was used by \cite{KuziodeNaray2004} to correct \HB{} fluxes for the stellar absorption in a sample of six LSB galaxies and is motivated by the careful analysis of \HII{} regions in HSB galaxies by \cite{McCall1985} and \cite{Oey1993}. Additionally, this choice for our data is independently supported by the agreement between the \HD{}/\HB{} and \HE{}/\HB{} ratios (see Section \ref{sec:confirmation}).

To visually represent the distribution of emission-line luminosity with respect to the continuum emission of UGC~628, we present in Figure~\ref{fig:spitzerRGB}(b) an \HB{} image of UGC~628 with the stellar continuum overlayed as isophotes.

\subsection{Region Identification and Diffuse Emission}
\label{sec:regions}
\begin{figure}
\begin{center}
  \includegraphics[width=\linewidth]{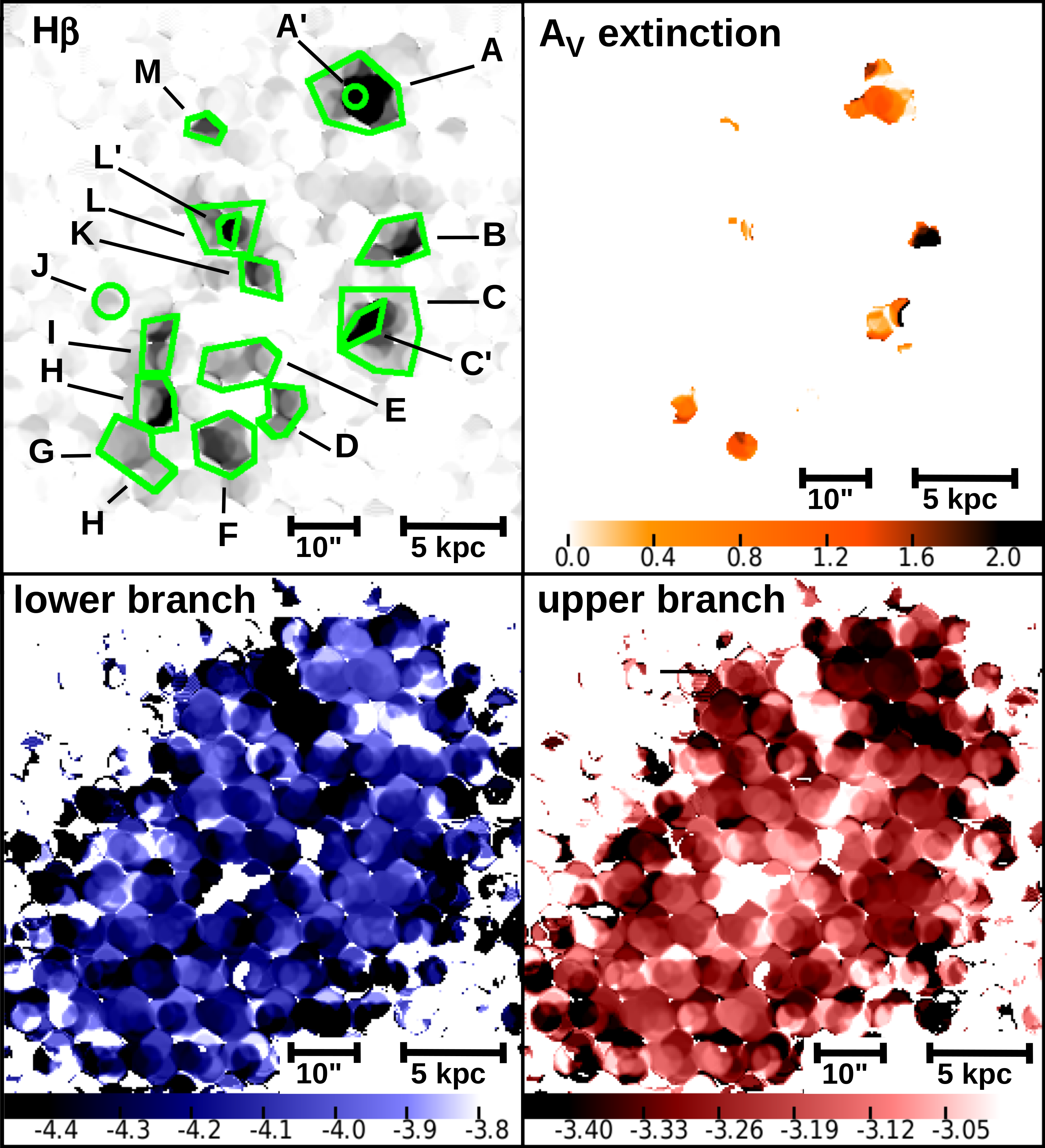}
  \caption{
    {\bf(a, top left)} \HB{} only, with \HII{} regions identified and labeled. Primed regions are particularly bright knots within larger regions.  {\bf (b, top right)} An A$_{\rm V}$ extinction map of UGC~628; note that the map is incomplete, as line emission is only detected from some areas. {\bf (c, bottom left)} A log(O/H) metallicity map of UGC~628 derived using the $R_{23}$ method adopting the lower-branch value (see Section \ref{sec:derived}).  {\bf (d, bottom right)} The complementary upper branch log(O/H) metallicity map of UGC~628. }
\label{fig:dustmetal}
\end{center}
\end{figure}

Visual inspection of Figure~\ref{fig:spitzerRGB}(b) strikingly reveals \HII{} regions arranged along the spiral arms of UGC~628. To analyze the sites of star formation as distinct entities we identified 16 discrete regions by visual inspection, labeled $A$-$K$ in Figure~\ref{fig:dustmetal}(a). These regions all met the criteria that they have centers 5$\sigma$ above the background, are roughly symmetric, and are separated from the next nearest region by a local minimum.

Linking these regions to canonical \HII{} regions is difficult because the 2\arcsec{} seeing at Mt. Locke corresponds to a physical scale of 770pc at the distance of UGC~628. Complicating matters further, the physical size of \HII{} regions is a thorny issue since no universally recognized method for delineating the edges of \HII{} regions exists. Clearly, our data cannot resolve small \HII{} regions such as the Orion Nebula, and the \HB{} bright regions in Figure~\ref{fig:dustmetal}(a) should be thought of as \HII{} region complexes and not individual \HII{} regions. However, three regions, \changed{$A$, $C$, and $L$}, have very bright centers with sizes near our 770pc limit, which are much more likely to be genuine giant \HII{} regions; we have identified these bright sub-regions as \changed{$A^\prime$, $C^\prime$, and $L^\prime$}. The properties of these regions, described below, are summarized in Table~\ref{tbl:HII}, along with the diffuse \HB{} emission and the net galaxy-wide \HB{} emission.

\subsection{Extinction Maps}
\label{sec:extinction}
To estimate the effects of extinction due to dust on the optical emission lines, we employed the \HB{} and \HG{} images and the extinction law described in \cite{Cardelli}. We have restricted our extinction calculations to regions where both the \HB{} and \HG{} lines are 3$\sigma$ above the threshold (see Section \ref{sec:weave}); this limits our analysis of the extinction in UGC~628 to only the emission line regions. 

We assumed an R$_{\rm V}$ of 3.1, as \cite{Cardelli} find this to be typical of the Milky Way, and an intrinsic \HG{}/\HB{} ratio of 0.466, characteristic of Case B recombination with $T\!=\!10,000\rm K$ \citep{Brocklehurst}. This information, along with our measured \HG{}/\HB{} map, allowed us to create $\rm A_\lambda$ maps for each of our observed emission lines; as an example, we present the $\rm A_V$ map in Figure~\ref{fig:dustmetal}(b).

\subsection{Confirmation}
\label{sec:confirmation}

As described in Section \ref{sec:calib}, we confirmed the spectrophotometric calibration by comparing our data to established broad-band photometry \citep{Kim2007}; furthermore, with the emission-line images in hand, we performed two additional tests which confirm our calibration as well as our continuum subtraction.

First, we compared the strength of the \ionl{O}{3}{4959} and \ionl{O}{3}{5007} lines. The transition probabilities give a theoretical ratio of \ionl{O}{3}{5007}/\ionl{O}{3}{4959}~=~2.98 \citep{Storey2000} and, since these lines are close in wavelength, this ratio is not strongly affected by reddening. We measure a ratio 3.11$\pm$0.04 in the 2\arcsec{} aperture centered on Region $A^\prime$ in the narrow-band images centered on \ionl{O}{3}{4959} and \ionl{O}{3}{5007}.

Second, the emission lines in Region $A^\prime$ are strong enough for us to detect the \HD{} and, marginally, the \HE{} line, giving another set of fixed line ratios. Using the extinction curve calculated from the \HB{} and \HG{} fluxes, we de-extinguished the \HB{} flux measured in Region $A^\prime$, transformed it to the appropriate values for \HD{} and \HE{} using the Case B $T\!=\!10,000\rm K$ ratios of 0.256 and 0.158, and finally re-extinguished them using the same extinction curve. The values we calculated in this fashion are $1.0\pm0.3\times10^{-16}$ and $5.8\pm0.2\times10^{-17}\rm erg\,s^{-1} cm^{-2}$ for \HD{} and \HE{}, respectively; the values we measured are $1.2\pm0.4\times10^{-16}$ and  $9.4\pm4.3\times10^{-17}\rm erg\,s^{-1} cm^{-2}$. We treat this excellent agreement with the \HD{} line and the acceptable agreement with the marginally detected \HE{} line as a confirmation of our data processing and choice of extinction curve.

\begin{table*}
\caption{Properties of \HII{} Regions}
\begin{tabular}{lccccccc}
\hline
Region 
&R$_{gal}$\footnotemark[1]
&B-V\footnotemark[2]
&log(L$_{\rm H\beta}$)\footnotemark[3]
&SFR$_{\rm H\beta}$
&A$_{\rm V}$
&\multicolumn{2}{c}{\underline{log(O/H)}}
\\
&(kpc)
&${\rm \left(mag           \right)}$
&${\rm \left(erg\ s^{-1}   \right)}$
&${\rm \left(M_\odot\ yr^{-1}\right)}$
&${\rm \left(mag           \right)}$
&lower
&upper
\\
\hline

A          &  13.5  &  0.641    &  39.4   &  0.10    $\pm$  0.06    &  0.902  &  -4.0$\pm$    0.11   &  -3.4 $\pm$   0.040   \\
A$\prime$  &  13.7  &  0.740    &  38.6   &  0.026   $\pm$  0.016    &  1.19   &  -4.0$\pm$    0.10   &  -3.4 $\pm$   0.038   \\
B          &  11.4  &  0.535    &  38.7   &  0.048   $\pm$  0.030    &  1.62   &  -4.1$\pm$    0.15   &  -3.3 $\pm$   0.061   \\
C          &  12.9  &  0.587    &  39.1   &  0.062   $\pm$  0.039    &  1.03   &  -4.1$\pm$    0.13   &  -3.3 $\pm$   0.052   \\
C$\prime$  &  10.7  &  0.695    &  38.6   &  0.013   $\pm$  0.0082  &  0.740  &  -4.1$\pm$    0.12   &  -3.3 $\pm$   0.047   \\
D          &  9.28  &  ...      &  38.5   &  0.0033  $\pm$  0.0020   &  ...    &  -4.1$\pm$    0.066  &  -3.3 $\pm$   0.026 \\
E          &  3.84  &  0.605    &  38.8   &  0.016   $\pm$  0.0099  &  0.296  &  -4.3$\pm$    0.11   &  -3.1 $\pm$   0.057   \\
F          &  9.14  &  -0.0855  &  38.8   &  0.016   $\pm$  0.0010   &  0.439  &  -4.2$\pm$    0.12  &  -3.2 $\pm$   0.055   \\
G          &  10.0  &  0.618    &  38.7   &  0.0071  $\pm$  0.0045   &  ...    &  -4.2$\pm$    0.11   &  -3.2 $\pm$   0.050   \\
H          &  7.60  &  ...      &  38.7   &  0.012   $\pm$  0.0075   &  0.376  &  -4.3$\pm$    0.11   &  -3.2 $\pm$   0.049   \\
I          &  7.08  &  ...      &  38.6   &  0.036   $\pm$  0.022    &  1.55   &  -4.2$\pm$    0.16   &  -3.2 $\pm$   0.067   \\
J          &  12.5  &  ...      &  38.00  &  0.0011  $\pm$  0.00068  &  ...    &  -4.1$\pm$    0.12   &  -3.3 $\pm$   0.055  \\
K          &  2.09  &  ...      &  38.6   &  0.026   $\pm$  0.017    &  1.25   &  -4.3$\pm$    0.15   &  -3.1 $\pm$   0.076   \\
L          &  7.07  &  0.513    &  38.9   &  0.023   $\pm$  0.014    &  0.582  &  -4.1$\pm$    0.13   &  -3.2 $\pm$   0.059 \\
L$\prime$  &  7.08  &  0.463    &  38.3   &  0.0053  $\pm$  0.0033   &  0.527  &  -4.2$\pm$    0.12  &  -3.2 $\pm$   0.054  \\
M          &  15.7  &  0.0965   &  38.2   &  0.0023  $\pm$  0.0015   &  ...    &  -4.0$\pm$    0.071  &  -3.4 $\pm$   0.027   \\

\hline
whole      &  ...   &  0.64     & 40.3   &  0.45 $\pm$ 0.028   &  ...    &  -4.2$\pm$0.020   &  -3.2   $\pm$  0.011  \\
diffuse    &  ...   &  ...      & 40.0   &  0.17           &  ...    &  ...               &  ...   \\
\hline\\\\

\multicolumn{2}{l}{$^1$ deprojected distance}\\
\multicolumn{8}{l}{$^2$ regions without B-V showed no statistically significant optical enhancement above neighboring pixels}\\
\multicolumn{5}{l}{$^3$ uncertainty dominated by 0.07mag uncertainty in photometric calibration}

\label{tbl:HII}
\end{tabular}
\end{table*}

\subsection{Optically Derived Properties}
\label{sec:derived}
Table~\ref{tbl:HII} lists the properties of the emission-line regions. The \HB{} luminosity was determined via extended source photometry in the \HB{} image (see Section \ref{sec:lineimages}); the star-formation rates were calculated from the extinction corrected \HB{} luminosity via the SFR/\HA{} in \cite{Kennicutt2009} with an \HA{}/\HB{} of 2.87, again assuming Case B recombination with $T=10,000$ K \citep{Brocklehurst}.

The B-V colors of the emission line regions in Table~\ref{tbl:HII} are the colors of the local optical enhancement in the $x$-$y$-$\lambda$ spectral data cube (referenced against Bessel B and V filters). The local B and V background for each region was calculated as the average value of the pixels falling on the perimeter of each region. All 16 \HB{} bright regions show at least some some optical enhancement, typically much bluer than the surrounding areas.

Table~\ref{tbl:HII} also lists log(O/H) metallicities calculated via the $R_{23}$ quantity derived from the \ionl{O}{2}{3727}, \HB{}, \ionl{O}{3}{4959}, and \ionl{O}{3}{5007} lines using the analytic formulae in \cite{KuziodeNaray2004}. The $R_{23}$ quantity is a bi-valued function of log(O/H) metallicity; for low metallicities, increasing metallicity increases the strength of oxygen lines; for high metallicities, increasing metallicity decreases the strength of the oxygen lines due to increased cooling via infrared lines.

Without the \ionl{N}{2}{6584} line to enable discrimination between the high metallicity `upper' branch and the low-metallicity `lower' branch, we present both metallicity values in Table~\ref{tbl:HII}.

\section{Results} 
\label{sec:results}

\subsection{Distribution of Star Formation}
\label{sec:sizelum}
The galaxy-wide star-formation rate measured via the extinction corrected \HB{} flux is $0.45\pm0.028$ $\rm M_\odot yr^{-1}$.  As discussed in Section \ref{sec:derived}, we calculated this value from the extinction corrected \HB{} luminosity using the case B recombination \HA{}/\HB{} ratio of 2.87 and the star-formation rate/\HA{} luminosity relationship from \cite{Kennicutt2012}. This star-formation rate is slightly above the average of the LSB galaxies in \cite{Kim2007} but well below the Milky Way's 1.9 $\rm M_\odot yr^{-1}$ \citep{Kennicutt2012}. Since \HA{} luminosity is the de facto standard when comparing star-formation rates between galaxies, we will make extensive comparisons between the \HA{} luminosities we derived from our \HB{} fluxes and the \HA{} luminosities available for other galaxies in the literature.

The most striking aspects of Figure~\ref{fig:spitzerRGB}(b), Figure~\ref{fig:dustmetal}(a), and Table~\ref{tbl:HII} are 1) the distribution of star formation in UGC~628 into discrete clumps and 2) the alignment of these clumps with the bar and spiral arms. While our 770pc resolution limits our ability to resolve the smallest \HII{} regions, we can say that the \HB{} bright regions that we have identified account for 50\% of the observed \HB{} luminosity but only 15\% of the area out to the same deprojected galactocentric distance (adopting the 56\deg{} inclination from \cite{Kim2007}). Star formation in UGC~628 appears to be extremely concentrated, {\it though noticeably absent in the galactic center.}

\begin{figure}
\begin{center}
  \hbox{
    \includegraphics[width=0.25\linewidth]{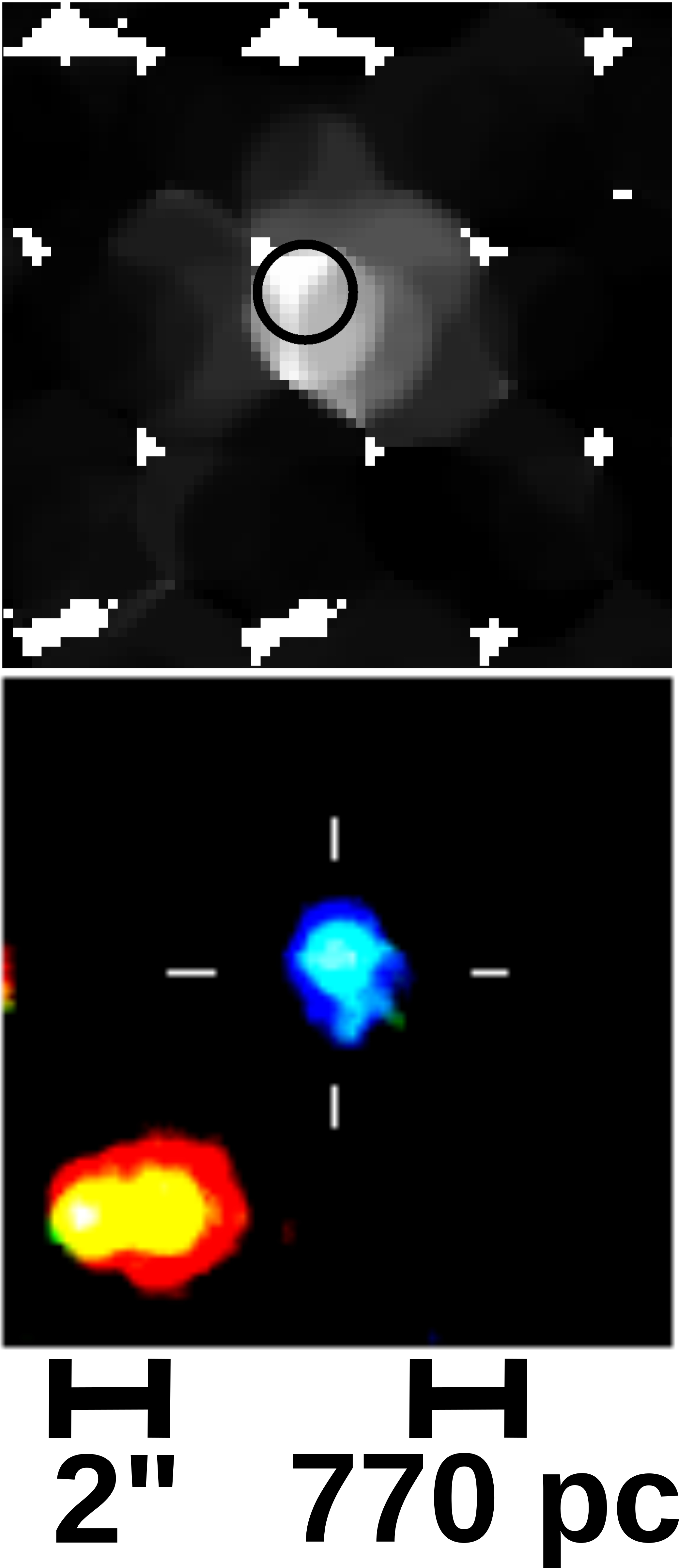}
    \includegraphics[width=0.75\linewidth]{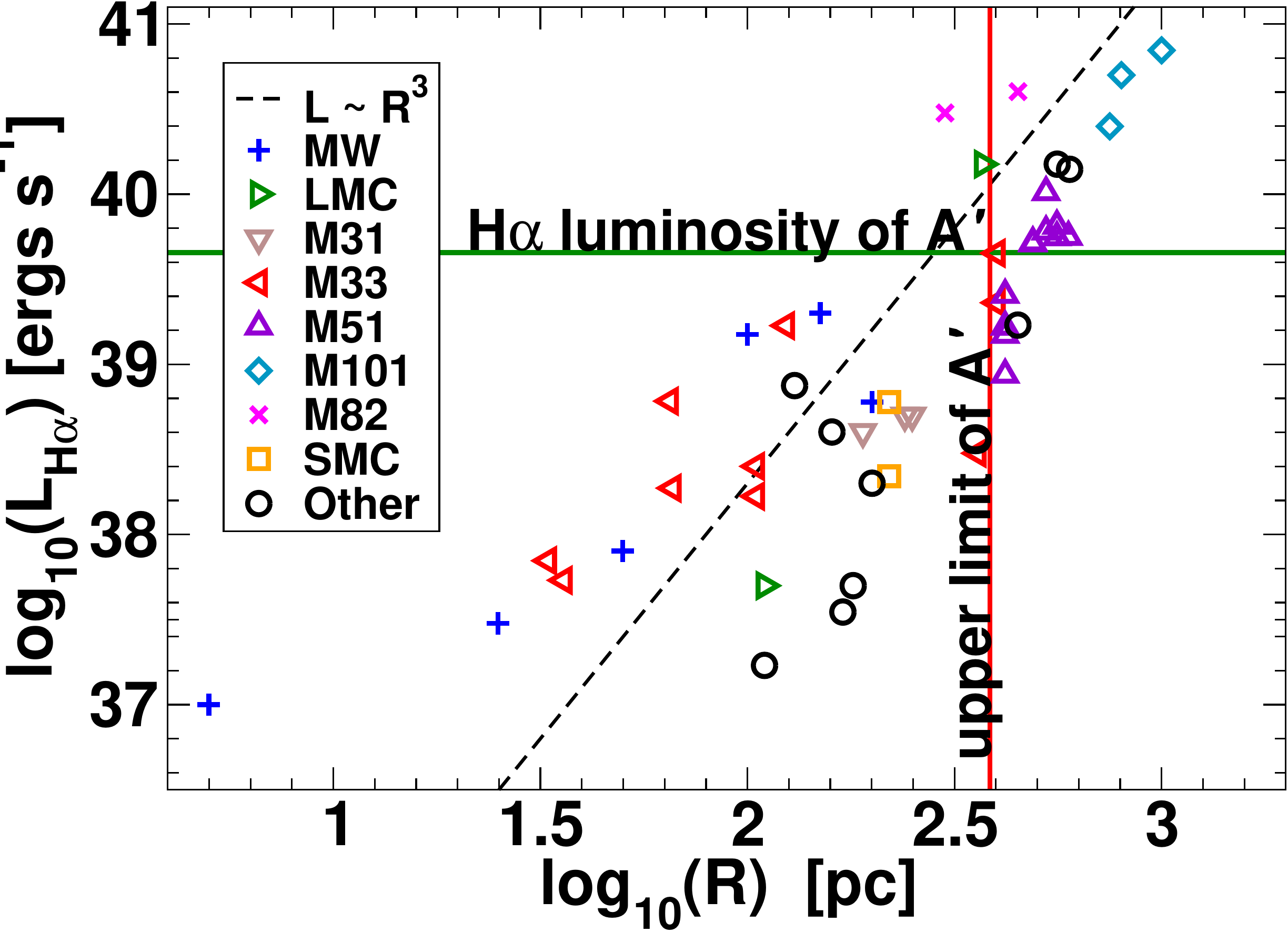}
  }
  \caption{
    {\bf(a, top left)} An enlargement of Figure~\ref{fig:dustmetal}(a) highlighting Region $A^\prime$. Region $A^\prime$ is defined by the black 2\arcsec{} circle, which has been centered to optimize the flux captured from this somewhat lopsided distribution. Note that the \HB{} enhancement is compact at the scale of 2\arcsec{}, the seeing limit of Mt. Locke. {\bf (b, bottom left)} An enlargement of the SDSS image in Figure~\ref{fig:spitzerRGB}(a) identifying Region $A^\prime$. {\bf (c, right)} A comparison of the diameter and unobscured \HA{} luminosities of \HII{} regions in nearby galaxies; marked on the plot is the calculated unobscured \HA{} luminosity of Region $A^\prime$, the seeing-imposed upper limit on the size of Region $A^\prime$, and, to guide the eye, an $\rm L_{H\alpha} \propto R^3$ line. Since the \HA{} luminosity of Region $A^\prime$ is in line with the upper limit on the diameter of Region $A^\prime$, it is highly likely to be a genuine giant \HII{} region. }
\label{fig:a-prime}
\end{center}
\end{figure}

Indeed, Region $A$, by itself, accounts for 20\% of the star formation in this galaxy, and Region $A^\prime$ accounts for 20\% of the star formation in Region $A$. Region $A^\prime$ is particularly interesting because it is concentrated even at the 2\arcsec{} scale testable by the VIRUS-P IFU (see Figure~\ref{fig:a-prime}(a)).

To place Region $A^\prime$ in context, we compare in Figure~\ref{fig:a-prime}(c) the calculated extinction-corrected \HA{} luminosity and the upper limit of the size of Region $A^\prime$ to \HII{} regions in nearby galaxies (M33, M31, LMC, SMC, M82, M101, M51) as well as in the Milky Way (M42, M8, Carina, W49) drawn from previous works \citep{Kennicutt1984,Viallefond1986, vanderHulst1988}. To guide the eye, we have also plotted an $\rm L_{H\alpha} \propto R^3$ line.

The fact that so many of the \HII{} regions from nearby galaxies fall near the $\rm L_{H\alpha} \propto R^3$ line partly reflects an inherent size/luminosity relationship, but also reflects the fact that divisions between \HII{} regions are often somewhat arbitrary. Many \HII{} regions which can be classified as single structures can just as easily be subdivided into smaller structures, often down to the level of individual Str\"{o}mgren spheres. The luminosity and size upper limit intersection of Region $A^\prime$ falls very close to this relationship, implying that it is likely either a single \HII{} region with an unobscured $\rm log(L_{H\alpha})\sim39.6$ or an agglomeration of several \HII{} regions immediately adjacent to each other. Since the divisions between immediately adjacent \HII{} regions are fairly arbitrary, these two scenarios are not meaningfully different in terms of size and intrinsic \HA{} luminosity.

Region $A^\prime$, then, is likely a giant \HII{} region with a star-formation rate almost twice that of the largest \HII{} regions in the Milky Way; while such a region appears extraordinary in an LSB galaxy with a global star-formation rate more than three times lower than that of the Milky Way \citep{Kennicutt2012}, it is quite in line with the late-type (Sm) morphology of UGC~628.

The link between the morphology of a galaxy and its \HII{} region luminosity function is well addressed by \cite{Caldwell1991}, who carefully reconstruct the luminosity functions of \HII{} regions in several nearby early-type (Sa) spirals and compare their findings to the similarly constructed luminosity functions of late-type spirals described in \cite{KEH1989}. One of the major conclusions drawn by \cite{Caldwell1991} is that star formation in early-type spirals is much more evenly distributed than in late-type spirals, and that the slope of the high end of the \HII{} region luminosity function is much steeper in early-type spirals. Normalizing for differences in global star-formation rate, early-type spirals are much less likely to have bright star forming regions ($\rm log(L_{H\alpha})\sim39$) than late-type spirals.

The strong clustering of star formation in UGC~628, highlighted by the presence of Region $A^\prime$, implies a shallow, top-heavy \HII{} region luminosity function consistent with other late-type spirals. While the argument is less compelling that Regions \changed{$C^\prime$ and $L^\prime$} are single \HII{} regions, the presence of such highly concentrated star formation supports the conclusion that UGC~628 has a very shallow \HII{} region luminosity function.

The \HII{} region luminosity function would seem, then, to be much more closely linked with morphological type than it is with global star-formation rate since the global star-formation rate of UGC~628 is quite low for a late-type spiral. This is in line with the findings of \cite{Schombert2013}, where they construct an \HII{} region luminosity function using 429 \HII{} regions in 54 LSB galaxies (mostly irregulars), and find that it matches the \HII{} region luminosity function for HSB irregulars presented in \cite{Youngblood1999}. Indeed, \cite{Schombert2013} compare net galaxy \HA{} luminosities to the \HA{} luminosities of the brightest \HII{} regions within each galaxy, and UGC~628 and Region $A^\prime$ would fit squarely within their data.

That said, UGC~628 and the LSB galaxies in \cite{Schombert2013} both share one significant difference when compared to many HSB spirals: a lack of detectable star formation in the core. \cite{Schombert2013} demonstrate with their sample of 429 \HII{} regions in LSB galaxies that the centers of LSB galaxies are generally devoid of \HII{} regions and suggest that this may be an indicator that star formation in LSB galaxies is driven by local fluctuations in gas densities rather than global patterns.  Nonetheless, star formation in UGC~628 is clearly influenced by some global process, as indicated by the strong alignment of emission-line regions with the bar and arms.  Indeed, arms and bars appear to be the sites of star formation in LSB galaxies just as they are in HSB galaxies since \cite{Kim2007} finds that nearly half of the spiral LSB galaxies in his sample show alignment of \HII{} regions with bars and/or spiral arms.

\begin{figure}
\begin{center}
  \includegraphics[width=\linewidth]{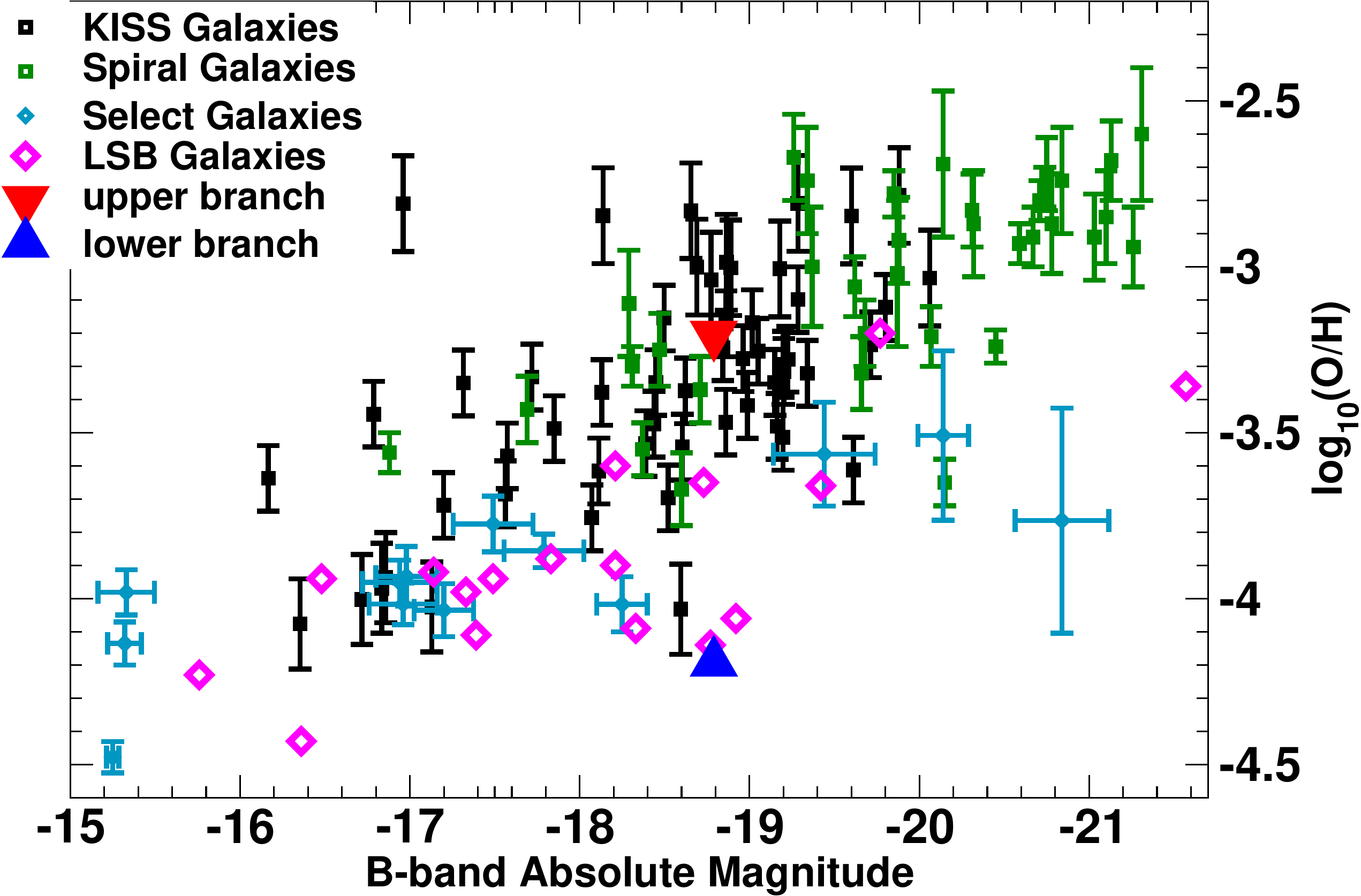}
  \caption[\ref*{Denicolor2002,KuziodeNaray2004,Young2014}]{
   \changed{Galaxy-averaged log(O/H) metallicity \vs{} B-band absolute magnitude relationship in UGC~628 (upper and lower branches) and a sample of LSB galaxies taken from \cite{KuziodeNaray2004} compared with that found in two samples of predominantly HSB galaxies: volume-limited \HA{} selected galaxies from the KISS project \citep{Young2014}, and spirals galaxies from \cite{Zaritsky1994}, as well as a diverse sample of galaxies from \cite{Denicolo2002}}. Both upper and lower branch metallicities are shown for UGC~628 (see Section \ref{sec:derived}). The upper branch metallicity falls well within the locus of spiral galaxies and volume-limited galaxies, while the lower branch value is more in-line with other LSB galaxies and similar to the metallicities of several dwarf galaxies from \cite{Denicolo2002}.

  }
\label{fig:kisscompare}
\end{center}
\end{figure}

\begin{figure*}
\begin{center}
  \hbox{
    \includegraphics[width=0.486\linewidth]{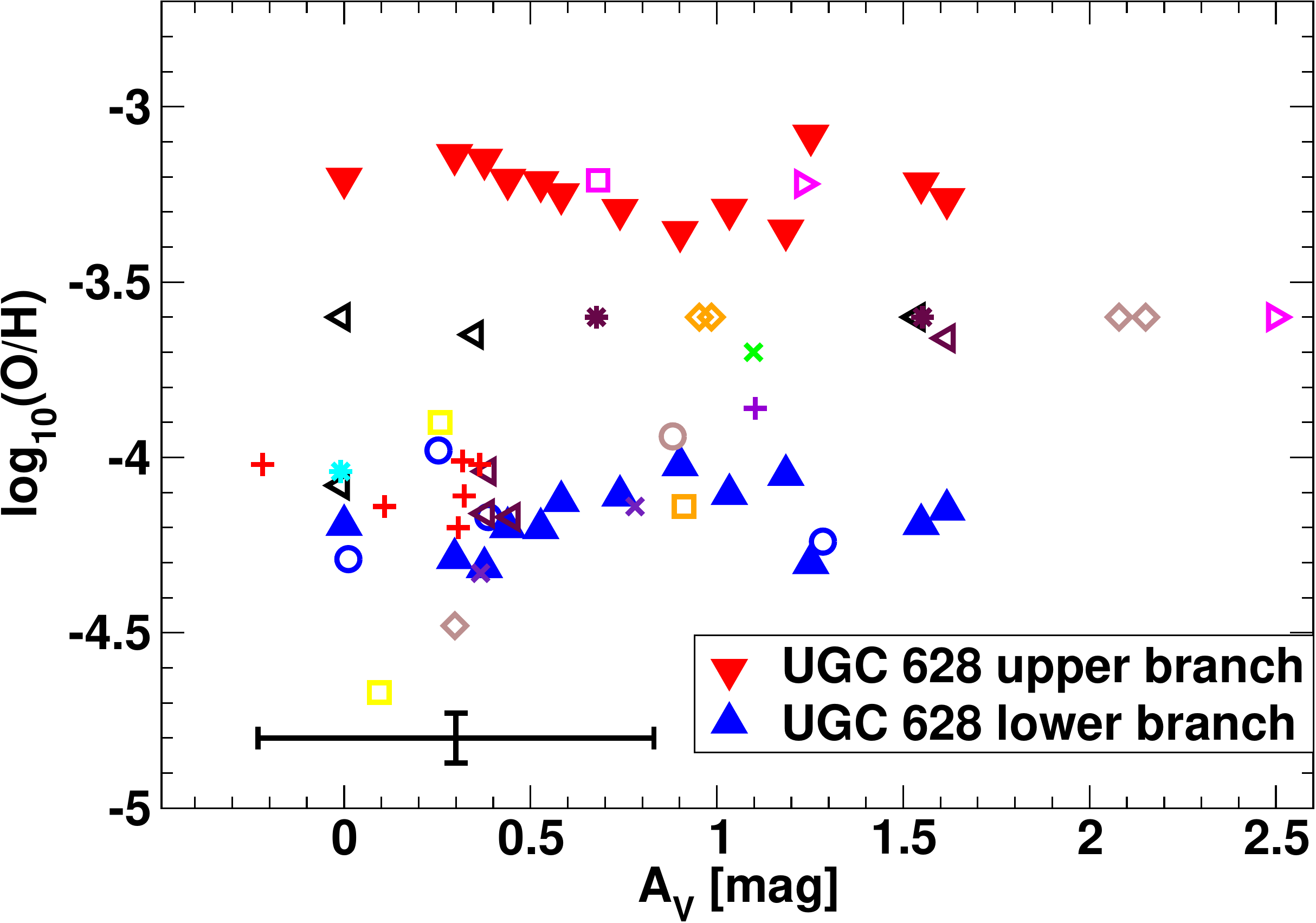}
    \includegraphics[width=0.5\linewidth]{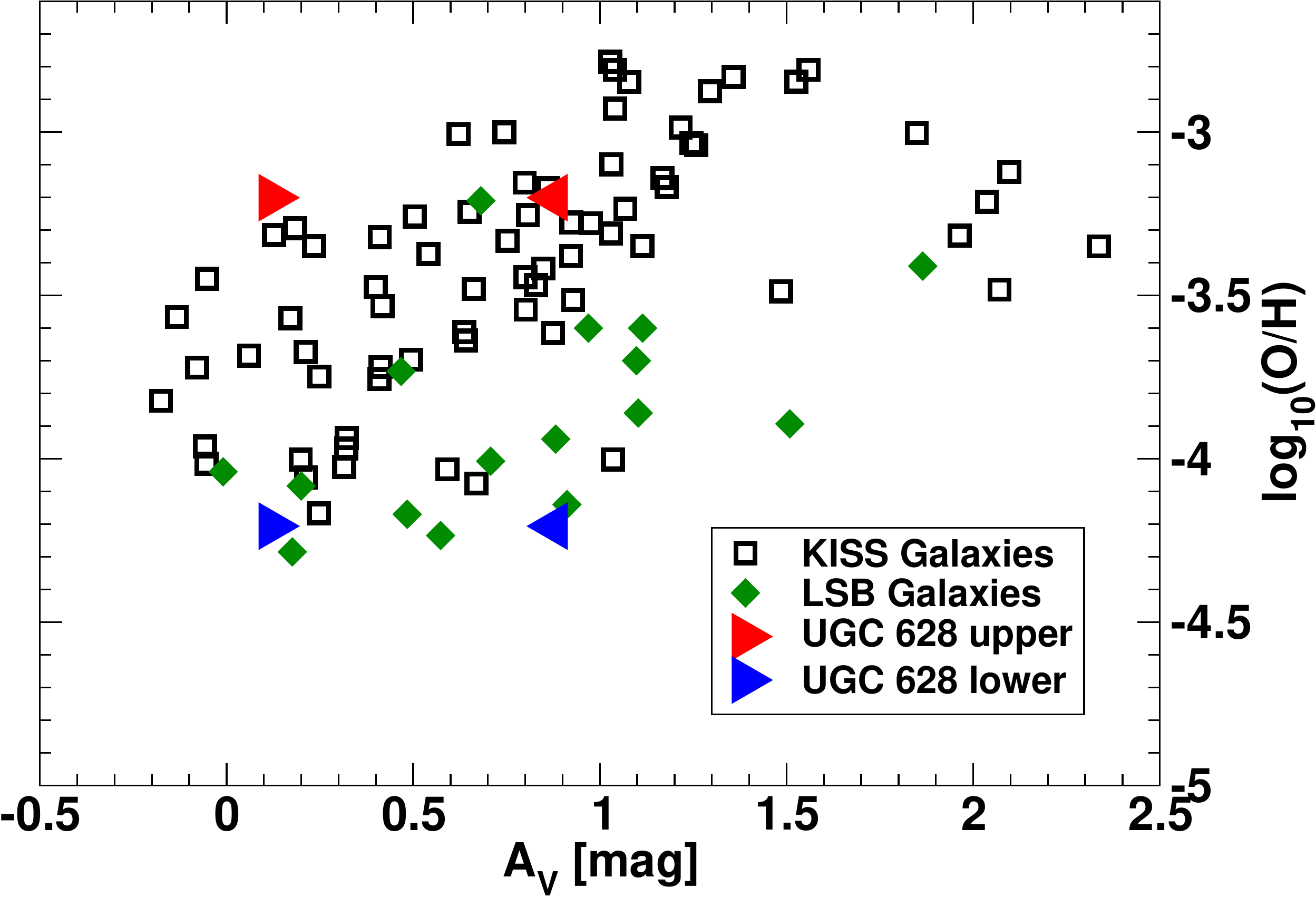}
   }
  \caption[\ref*{McGaugh1994abundance,Young2014,McGaugh1994abundance}]{\
    {\bf (a, left)} A comparison of log(O/H) metallicity and \AV{} extinction for individual \HII{} regions in UGC~628 and the LSB galaxies reported in \cite{McGaugh1994abundance}. Both upper and lower branch metallicities are plotted for UGC~628. For the \HII{} regions from \cite{McGaugh1994abundance}, each host galaxy is coded with a distinct symbol; \HII{} regions with identical symbols are from the same galaxy. Typical errorbars for UGC~628 are shown in the lower left.  {\bf (b, right)} Galaxy-averaged log(O/H) metallicity \vs{} \AV{} extinction for UGC~628, \HA{} selected galaxies from the KISS sample in \cite{Young2014}, and LSB galaxies from \cite{McGaugh1994abundance}. Both upper and lower branch metallicities are shown for UGC~628 (see Section \ref{sec:branches}). Since we only measure extinction in regions of UGC~628 where \HB{} and \HG{} are detected, the \AV{} values plotted are upper and lower limits. The upper limit is a flux-weighted average {\it excluding} regions where \HB{} or \HG{} were undetected, and the lower limit is a flux-weighted average including all areas where \HB{} was detected, assuming zero extinction in the absence of \HG{}. Similarly, the \AV{} values reported in \cite{McGaugh1994abundance} are for \HII{} regions, and should be taken as upper-limits on the galaxy-averaged extinction.}

\label{fig:metaldust}
\end{center}
\end{figure*}

\subsection{Upper vs. Lower Branch Metallicities}
\label{sec:branches}
While absolute metallicities cannot be directly determined using the $R_{23}$ strong-line method because of the ambiguity between the upper and lower branches, context clues favor the lower branch for UGC~628. For example, an examination of Table \ref{tbl:HII} shows that the most compact and vigorous star forming regions, $A$ and $A^\prime$, have the highest lower branch metallicities and the lowest upper branch metallicities. Since star forming regions are the sites of enrichment is highly unlikely that the most rapidly star forming regions in UGC~628 would also have the lowest metallicity. A similar argument can be made for Regions $B$, $C$, and $C^\prime$, the next most vigorous star forming regions in UGC~628, all of which have particularly high lower branch metallicities and low upper branch metallicities. The upper branch interpretation of $R_{23}$ is essentially impossible to reconcile with the relative star-formation rates of the specific emission line regions.

Additional context clues can be found by comparing the metallicity of UGC~628 to other galaxies. \changed{In Figure~\ref{fig:kisscompare} we compare the metallicity \vs{} B-band absolute magnitude relationship in UGC~628 (upper and lower branches) and a sample of LSB galaxies taken from \cite{KuziodeNaray2004} with that found in two samples of predominantly HSB galaxies: volume-limited \HA{} selected galaxies from the KISS project \citep{Young2014}, and spirals galaxies from \cite{Zaritsky1994}, as well as a diverse sample of galaxies from \cite{Denicolo2002}.} The well-established metallicity/luminosity relationship is clearly apparent, with LSB galaxies falling on the lower envelop. The galaxy-wide average upper branch metallicity of UGC~628, plotted as a red triangle, is similar to the metallicities of HSB spirals, while the lower branch metallicity, plotted as a blue triangle, is more consistent with other LSB spirals.

A comparison of metallicity to dust extinction also favors the lower branch interpretation. The relationship between dust and star formation is complex: dust is an indicator of the dense molecular clouds that host star formation, it plays a role in the cooling processes that determine the IMF, and it is comprised of the metals that result from the cumulative star-formation history. In Figure~\ref{fig:metaldust}(a) we plot log(O/H) metallicity \vs{} \AV{} extinction for the \HII{} regions in UGC~628 for which we have both measurements as well as \HII{} regions in other LSB galaxies described in \cite{McGaugh1994abundance}. Again, both upper and lower branch metallicities are plotted for UGC~628. The data show a wide scatter, but there is a conspicuous absence of points in the lower-right corner, consistent with the idea that metals are required for the formation of dust grains. 

The lower branch metallicities of UGC~628 fall within the primary locus of points in Figure~\ref{fig:metaldust}(a), while the upper branch metallicities fall at the very upper extreme of \HII{} regions in LSB galaxies. Indeed, if the upper branch metallicities were correct, that would imply that some \HII{} regions in UGC~628 would have solar or super-solar metallicities with nearly zero extinction.

In Figure~\ref{fig:metaldust}(b) we plot galaxy-averaged metallicity and extinction for UGC~628, the \HA{} selected galaxies from \cite{Young2014}, and the LSB galaxies from \cite{McGaugh1994abundance}. The metallicities and extinctions for the \HA{} selected galaxies were determined via long-slit spectroscopy, while the metallicities and extinctions for the LSB galaxies are an average of individual \HII{} regions within each galaxy. Again, we represent both the upper and lower branch metallicities for UGC~628. The galaxy-averaged extinction for UGC~628 was calculated in two ways. First, a direct \HB{} flux-weighted average of the \HII{} regions, and, second, an \HB{} flux-weighted average assuming an extinction of zero for areas that are detectable in \HB{} but too faint in \HG{} for a proper extinction calculation. The first method represents an upper limit on the true galaxy-wide average extinction since the regions which are brightest in emission lines are systematically more heavily extinguished. Likewise, the second method represents a solid lower limit on the galaxy-wide average extinction since the areas which are emission-line faint likely have at least some level of extinction, albeit systematically much lower than the emission-line bright regions. The true galaxy-averaged extinction falls between these two limits. Thus UGC~628 is represented as four points in this figure, two for the upper branch and two for the lower branch metallicities. The galaxy-averaged extinctions for the LSB galaxies from \cite{McGaugh1994abundance} have a similar ambiguity, however we are only able to plot the upper limits on extinction as data is only available for the individual \HII{} regions.

\changed{A correlation between metallicity and extinction is apparent among the primarily HSB KISS galaxies; the LSB galaxies also follow this correlation, but fall on the lower envelope.} The lower branch metallicity points for UGC~628 decidedly bracket the LSB galaxies. While the true galaxy-averaged extinctions of some of the LSB galaxies likely fall to the left of their plotted upper limits, the majority of these extinction values are likely not far from their plotted upper limits since the half-magnitude shift that would be needed to bring all of the extinction values in-line with the \HA{} selected galaxies would also lead to physically implausible negative extinction values for several galaxies.

Based on the context presented, it is our conclusion that the lower branch metallicities are much more favored than the upper branch metallicities.

\subsection{Metallicity Gradient}
\label{sec:gradient}

\begin{figure}
\begin{center}
  \hbox{
    \includegraphics[width=\linewidth]{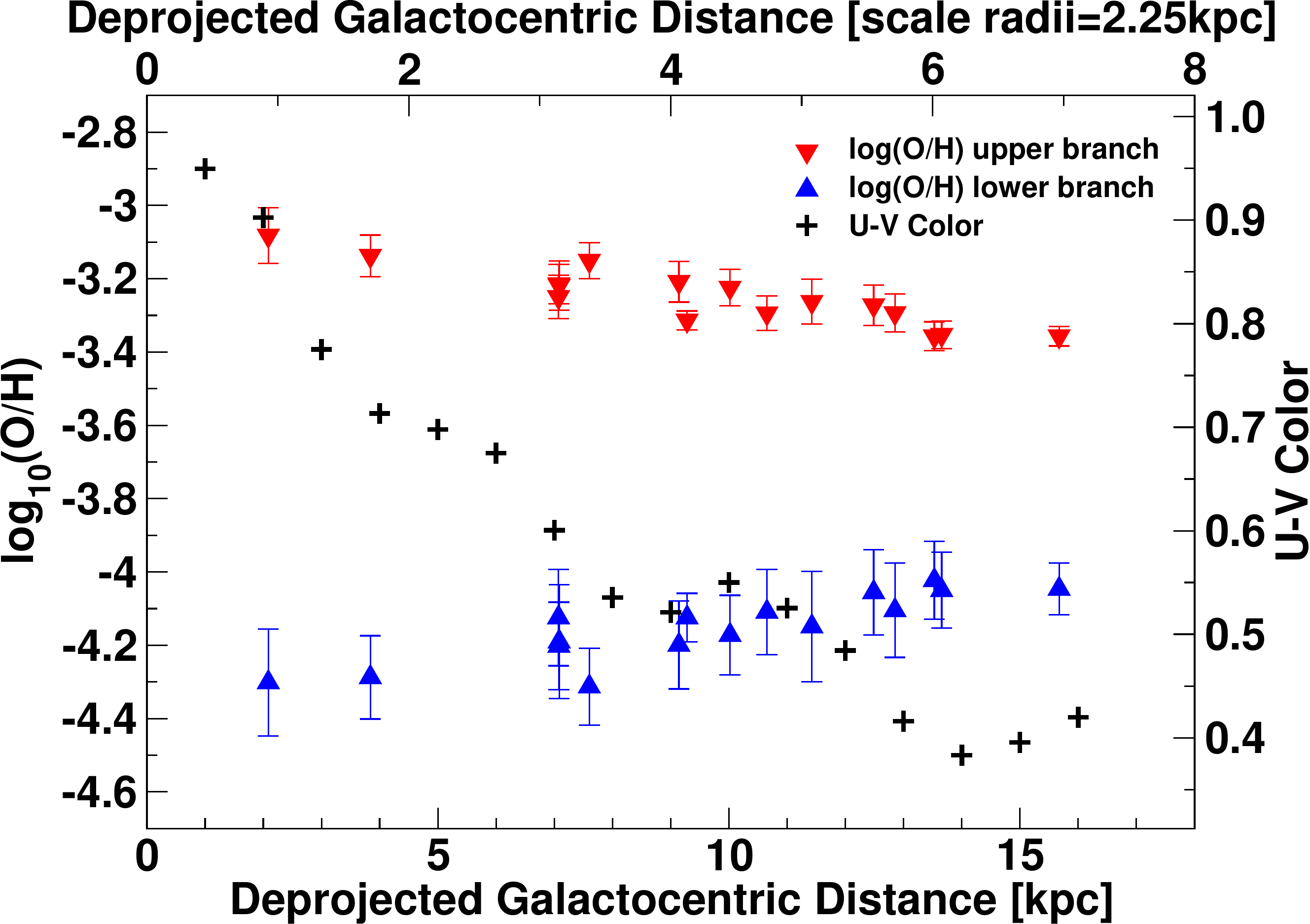}
   }
  \caption{log(O/H) metallicity of individual \HII{} regions within UGC~628 \vs{} deprojected galactocentric distance. On the same plot, U-V color of annuli at constant galactocentric distances. Both upper and lower branch metallicities are plotted (see Section \ref{sec:branches}); the upper branch shows a negative gradient and the lower branch shows a positive gradient while the U-V color shows a strong gradient of blue to red. }
\label{fig:gradient}
\end{center}
\end{figure}

In Figure~\ref{fig:dustmetal}(c,d) we show metallicity maps for both lower and upper branches, and in Figure~\ref{fig:gradient} we plot the average metallicity within each of the \HB{} bright regions listed in Table~\ref{tbl:HII} against their galactocentric radii. We find a monotonic slope in both branches; the upper branch has a slope of $-0.020\pm0.002\,\rm dex/kpc$ and the lower  branch has a slope of $+0.021\pm0.003\,\rm dex/kpc$.  The difference in signs for these slopes is related to the behavior of $R_{23}$ with metallicity, as discussed in Section \ref{sec:derived}. Nearly all low redshift HSB spirals show a negative metallicity gradient, typically $\sim\ $-0.05 dex/kpc, with higher metallicities near the centers \citep{Zaritsky1994,vanZee1998,Ho2015}. If we adopt the upper branch values for UGC~628, then this matches typical spiral galaxies quite well. If, however, we adopt the lower branch values, which are much more supported by the context described in Section \ref{sec:branches}, UGC~628 is fairly unusual for a spiral galaxy.

The prevailing explanation for the ubiquity of negative metallicity gradients is the inside-out hypothesis, wherein stars form earlier and faster in the central regions of galaxies \citep[e.g.,][]{Goetz1992}. This hypothesis is supported by observations of steeper metallicity gradients in spirals out to z$\sim$2 \citep{Jones2010,Yuan2011}, suggesting that metallicity gradients were steepest near the peak of cosmic star formation.

There are several significant exceptions to the ubiquity of negative metallicity gradients which also support this explanation. \cite{Kewley2010} find that close pairs of galaxies have shallower metallicity gradients, and conclude that gas tidally driven from the outskirts to the centers of interacting galaxies likely flattens their metallicity gradients. Similarly, \cite{Cresci2010} identify three z$\sim$3 galaxies with inverted metallicity gradients, and suggest that, at this early epoch, prior to the peak of star formation but during the era of mass assembly, primordial gas accretion onto galactic centers dominates.

LSB galaxies may constitute another significant exception to the ubiquity of negative metallicity gradients, and, thereby, provide another key insight into galaxy evolution. \cite{deBlok1998I} measure the metallicity profiles of three LSB galaxies using the $R_{23}$ method used here and the FIVEL method described in \cite{DeRobertis1987}, and are unable to find a metallicity gradient. The metallicity profile of UGC~628, however, is constrained much more tightly and out to more than twice the galactocentric scale radius because of the larger number of data points in our sample.

The distribution of star formation in UGC~628 hints that inside-out formation may not be an appropriate model for LSB galaxies. We we noted in Section \ref{sec:sizelum}, star formation in UGC~628 is decidedly absent in the core, with the large emission-line regions located in the outer spiral arms. A simple interpretation of the positive metallicity gradient is that UGC~628 has experienced outside-in formation, and that the outer disk is more metal rich due to a longer and/or more vigorous history of star formation. This explanation is supported by circumstantial evidence from other LSB galaxies; for example, \cite{Galaz2006} and \cite{Matthews1997} find inverted (red-to-blue) color gradients in 20\% and 50\% of LSB galaxies, respectively. Radial migration of material certainly plays a role in determining the metallicity gradient, but it is unlikely to be a large part of the explanation for LSB galaxies since many LSB galaxies are in nearly solid-body rotation\citep[e.g.,][]{deBlokMcGaugh2001a,deBlokMcGaugh2001b,deBlokBosma2002}, and are expected to be fairly inefficient at transporting material radially. If anything, they are likely to retain relic metallicity gradients should such a gradient ever develop.

A complexity arises when we consider the clear, monotonic negative U-V color gradient exhibited by UGC~628 (over-plotted in Figure~\ref{fig:gradient}). The negative color gradients seen in most HSB spirals are another line of evidence for inside-out formation; the negative color gradient in UGC~628 seems to signal inside-out formation, contrary to the metallicity gradient. One possible explanation for this apparent paradox is that star formation was more vigorous and more centrally concentrated in the past, but has since slowed and moved to the outer disk. This is in-line with comparisons between models and broad-band photometry \citep[e.g.,][]{Zackrisson2005} and equivalent widths \citep[e.g.,][]{Vorobyov}, which conclude that LSB galaxies, while blue, are many Gyr old. In this scenario the positive present-day metallicity gradient establishes an upper limit on the inside-out star-formation at earlier times, determined by the specifics of ISM mixing. From a practical standpoint, it is more likely that a rigorous measurement of the star-formation history of UGC~628 and other LSB galaxies with flat or positive metallicity profiles could help constrain ISM mixing models.

A significant line of evidence supporting this hypothesis comes from the fact that UGC~628 contains a bar, a fairly rare structure in LSB spirals \citep{Mihos1997}. The scarcity of bars in LSB galaxies is likely due to the fact that solid-body rotators are particularly stable against bar formation; however because of this stability, should a bar ever form, it would likely persist for many dynamical times. The bar in UGC~628 may be just such a relic since UGC~628 is almost in solid-body rotation out to $\sim$3 kpc \citep{deBlokBosma2002}, where the bar ends. If this is the case, the bar in UGC~628 signals that this galaxy was once more susceptible to global patterns than it is now, and may have exhibited inside-out formation at some point in the past.

If star formation in typical LSB galaxies favors a model other than inside-out formation, this would conveniently explain the reversed color gradients and flat metallicity profiles seen in other LSB galaxies. UGC~628, in this context, would be unusual for an LSB galaxy, having experienced a destabilizing event many dynamical times ago which left it with a relic bar and negative color gradient, but a metallicity profile that has been inverted by subsequent non-inside-out star formation. In the upcoming series of MUSCEL papers our group will present robust, spatially resolved star-formation histories needed to flesh out or refute such a scenario for UGC~628 as well as other LSB galaxies in our sample.

\subsection{Metallicity as a Tracer of Chemical Evolution}
The evidence presented in Section \ref{sec:branches} indicates that the lower branch metallicity is a more likely interpretation of our data. Throughout most of the disk of UGC~628 the metallicity is close to $\rm log(O/H)\sim-4.2$ (\rfrac{1}{6} solar), with the most metal rich \HII{} regions peaking around $\rm log(O/H)\sim-4.0$ (\rfrac{1}{4} solar). While this value is typical for dwarf galaxies, low metallicities such as this are rare in massive galaxies. For example, in Figure~\ref{fig:kisscompare} there are many HSB galaxies with metallicities at or below -4, but the majority of them are at least two absolute magnitudes less luminous than UGC~628. We can compare UGC~628 to HSB analogs by calculating what its metallicity would be if it followed the well-known mass-metallicity relationship. For the sake of example, if we adopt for UGC~628 $\rm log(M_\star/M_\odot) = 10.65$ from \cite{Kim2007} and the mass-metallicity coefficients from \cite{Tremonti2004}, we predict log(O/H)$=-2.9$, more than a dex greater than the -4.2 that we report. Clearly, UGC~628 deviates significantly from the mass-metallicity relationship that holds so well for HSB galaxies. As massive yet metal poor disks, LSB galaxies such as UGC~628  can best be placed in the context of galaxy evolution by considering the physical processes driving the metallicity, metallicity gradient, and the formation of dust.

The low metallicities of dwarf galaxies are generally attributed to their low escape velocities, which make them more susceptible to losing their ISM through supernova feedback \cite[e.g.,][]{Hunter1995,Thuan1999}. This has the dual effect of removing metal-enriched gas and also extinguishing star formation. However, unlike dwarfs, UGC~628 resides in a deep potential well that makes ISM blowout from supernova feedback an unlikely explanation. For example, \cite{Dekel2003} find that supernova feedback is no longer effective at evacuating the ISM from halos more massive than $3\times 10^{10}\rm M_\odot$; the rotation curve of UGC~628 indicates a dynamical mass of $6.5\times 10^{10}\rm M_\odot$ within 13.8~kpc. This key difference makes massive LSB galaxies particularly interesting for models of galaxy evolution because, with ISM removal ruled out, the low metallicity of UGC~628 and other LSB disks must be explained through the history of star-formation and gas accretion.

With this in mind, it is tempting to try to explain the low metallicities entirely through minimal past star-formation and think of UGC~628 and similar LSB galaxies as unevolved analogs to high redshift galaxies. Despite the great difficulty in measuring optical emission lines at $z\ga 2$ and the large intrinsic galaxy-to-galaxy variation, the trend of decreasing metallicity with increasing redshift is seen across a wide range of environments and galaxy types \citep[e.g.,][]{Savaglio2005,Erb2006,Rafelski2012,Bayliss2014,Sanders2015}. The standard interpretation here is an evolution of chemical enrichment over time. 

The analogy between high redshift galaxies and LSB galaxies is, however, imperfect at best, with a growing body of evidence suggesting that the stellar populations of LSB galaxies are quite old. For example, \cite{Zackrisson2005} study the star-formation histories of nine LSB galaxies using broad-band photometry and find that, while precise histories cannot be determined from only broad-band photometry, ages younger than 1.5~Gyr are categorically ruled out. Meanwhile, \cite{Vorobyov} compare \HA{} equivalent widths from a sample of LSB galaxies to a simulated LSB disk, and find that the LSB galaxies must be at least 5-6~Gyr old.  \cite{Vorobyov} also predict that a truly young LSB disk would exhibit strong radial fluctuations in metallicity due to local enrichment; quite to the contrary, the metallicity profile of UGC~628 is quite smooth and somewhat shallow. Indeed, the variation of only 0.3 dex across 10kpc, more than three scale lengths, categorically rules out scenarios relying entirely on recent enrichment and implies that most of the enrichment occurred many dynamical times ago. Clearly, LSB galaxies have undergone significant evolution, albeit along a different track than HSB galaxies.

A more complete view of chemical enrichment \citep[e.g.,][]{Lilly2013} treats the current metallicity as an equilibrium between infalling gas, star formation, and stellar feedback. Recently \cite{Peng2015} find that the chemical difference in the stellar populations of massive galaxies is best explained by a model where the ISM of star-forming galaxies is continuously diluted by infalling gas, while the ISM of quenched galaxies functions as a closed box. Once a galaxy's supply of infalling gas is removed, the metallicity of newly-formed stars rises rapidly until the gas is exhausted; this is in line with the consistently higher stellar metallicities that \cite{Peng2015} find in quenched \vs{} star-forming galaxies in their sample of 26,000 objects.

UGC~628 likely represents the corollary --- a galaxy where star formation slowed (or remained slow) despite a continued gas supply. The ultimate cause of the slow star formation remains unclear, but, regardless, if much of the stellar population of UGC~628 was produced many dynamical times ago then infalling pristine gas would have ample time to dilute the ISM metallicity and push UGC~628 off the mass-metallicity relationship. As discussed in Section \ref{sec:gradient}, the positive metallicity gradient along side the negative color gradient supports the thinking that UGC~628 is quite old, and enough time has passed for the metallicity gradient to be inverted by more recent (though possibly slower) outside-in star formation.

\changed{This scenario is in line with Figure~\ref{fig:metaldust}, where we see that, for a given metallicity, LSB galaxies typically have higher \AV{} extinctions than KISS galaxies, which are predominately HSB galaxies;} the ISM gas in LSB galaxies may have been diluted by infalling gas, but nothing has removed or destroyed the dust grains. Interestingly, \cite{Gerritsen1999} hypothesize that, with a low metal content, LSB galaxies have difficulty forming molecular clouds and become trapped in a low metallicity state. If this hypothesis is correct, newly formed galaxies in the early universe may have been effectively racing against infalling gas; those that failed to produce enough metals to keep the process going became trapped in the LSB state.

\section{Summary and Conclusions}
\label{sec:discussion}
We have utilized the emission line luminosities and ratios in UGC~628, the first target of our MUSCEL program, to determine the magnitude and distribution of star formation and gas-phase metallicity. Our results can be summarized as follows:

{\bf 1)} Star formation in UGC~628 is highly concentrated into a small number of large \HII{} regions, with very little star formation in the core. This tentatively supports the burst-and-quench model, where LSB galaxies are kept blue by brief but concentrated bursts of star formation which are unable to build up a substantial stellar populations.

{\bf 2)} The presence of the largest \HII{} regions rules out a steep \HII{} region luminosity function, making UGC~628 more consistent with other late-type galaxies even though the global star-formation rate is more typical of early-type galaxies. This implies that the mechanism which suppresses star formation in LSB galaxies arrests the process of \HII{} region formation early on, and that, once an \HII{} region begins to form, its growth proceeds as it normally would in a late-type galaxy.

{\bf 3)} The lower-branch $R_{23}$ metallicity is the most likely interpretation of the observed emission-line intensities; a typical value is log(O/H)$\sim$-4.2. This metallicity makes UGC~628 more metal poor than HSB spirals with similar masses but more heavily extinguished in \AV{} than HSB spirals with similar metallicities.

{\bf 4)} The metallicity shows a positive gradient across the galactic disk, with more metal rich areas nearer to the edge of the disk. This is at odds with a more typical color gradient (reddest in the center, bluer with larger radii), which implies that stellar populations near the center of UGC~628 have experienced greater cumulative star formation. It is likely that UGC~628 did experience inside-out formation long ago, and that the current outside-in mode of star formation has since inverted the metallicity gradient. This scenario demands that the stellar population near the center of UGC~628 is older than the characteristic time scale of the current star formation.

{\bf 5)} The low metallicity and high \AV{}/log(O/H) suggest that infalling pristine gas has diluted the metallicity without destroying the dust grains. This implies that much of the metal content in UGC~628 was produced at least several gas infall times ago.

These lines of evidence suggest that UGC~628 and similar LSB spirals may be the outcome of galactic `overfeeding'; that is, disks where continued star formation is stalled out by a low metallicity ISM created by a continuous supply of pristine gas when their initial burst of star formation is delayed, suppressed, or simply insufficient. This scenario, as well as others, will be rigorously tested in the upcoming series of papers where we will fit candidate star-formation histories against our optical spectra combined with UV and IR photometry for each region within UGC~628 and the remaining LSB galaxies in the MUSCEL sample.\\

\noindent{ACKNOWLEDGMENTS}

We thank Robin Ciardullo for many thoughtful insights into the planning, execution, and interpretation of our observations.

\bibliographystyle{apj}
\bibliography{lsb}

\end{document}